\documentclass[runningheads]{llncs}
\usepackage[T1]{fontenc}
\usepackage{graphicx}
\usepackage{xcolor,amsmath}
\begin{document}
\titlerunning{SAGE-FIN: A semi-supervised GNN approach with Causal Explanations} 
\title{
Detecting Fraud in Financial  Networks: \\ 
A Semi-Supervised GNN Approach\\
 with Granger-Causal Explanations  
}
%
%


\author{Linh Nguyen\inst{1}\orcidID{0009-0006-2797-5598}  \and Marcel Boersma\inst{1,2}\orcidID{0000-0001-9038-6020} \and Erman Acar\inst{1}\orcidID{0009-0002-9656-7249}}

\authorrunning{Nguyen et al.}
%
 \institute{University of Amsterdam, 1098 XH Amsterdam, the Netherlands \and
 KPMG Netherlands, 1186 DS Amstelveen, the Netherlands}
\maketitle              
\begin{abstract}

Fraudulent activity in the financial industry costs billions annually. Detecting fraud, therefore, is an essential yet technically challenging task that requires carefully analyzing large volumes of data. While machine learning (ML) approaches seem like a viable solution, applying them successfully is not so easy due to two main challenges:  (1) the sparsely labeled data, which makes the training of such approaches challenging (with inherent labeling costs),  and  (2)   lack of explainability for the flagged items posed by the opacity of ML models,  that is often required by business regulations. This article proposes SAGE-FIN, a semi-supervised graph neural network (GNN) based approach with Granger causal explanations for Financial Interaction Networks. SAGE-FIN learns to flag fraudulent items based on weakly labeled (or unlabelled) data points. To adhere to regulatory requirements, the flagged items are explained by highlighting related items in the network using Granger causality. We empirically validate the favorable performance of SAGE-FIN on a real-world dataset, Bipartite Edge-And-Node Attributed financial network (Elliptic++), with Granger-causal explanations for the identified fraudulent items without any prior assumption on the network structure.

\end{abstract}
\section{Introduction}
Fraudulent activities in transaction data present a critical yet challenging issue within the financial industry. Financial fraud arises in various areas, such as insurance, banking, taxation, and corporate sectors \cite{Albashrawi2016,Ngai2011}. Mitigating efforts are often labor intensive, requiring significant time, effort, and resources, costing billions annually; for example, Know-Your-Customer (KYC) departments within banks are estimated to cost 500 million dollars each year per bank. Machine learning (ML) approaches could alleviate this labor-intensive task, analyzing many data points for fraud characteristics. However, there remain two main challenges to tackle: (1) the labels of the financial transactions (fraud or non-fraud) are often expensive to obtain for model training~\cite{Chandola2009,Hu2019,Monamo2016,visbeek2024explainable}, and (2) the outcomes of the ML models often lack interpretability when it comes to justifying the detected anomalous items for further investigation. Motivated by these challenges, we propose a model that leverages the sparsely labeled data points and explains why the detected activities are flagged as fraudulent. 

Transactions can naturally be represented as graphs whose nodes correspond to involved parties and the edges correspond to money flows between them \cite{Acar2024,Boersma2024,Boersma2018}. The model must incorporate the graph structure to detect novel fraudulent patterns. Graph Neural Networks (GNNs) offer a framework for analyzing such structures and detecting novel patterns. Despite their demonstrated efficiency, most GNNs function as a black box without explicit knowledge representations, limiting their capability to provide necessary explanatory medium. This limitation, in turn, undermines their adoption in critical domains such as the financial sector~\cite{Bussmann2020-ns}. Rather necessary is the transparency of models to ensure traceable decision-making processes~\cite{Bussmann2020-ns}, comply with legal regulations~\cite{Gomber2018-hh} to maintain accountability while adhering to privacy~\cite{Weber2024-li}.
A desirable (informal) property of these explanations 
is that the generated explanations should reflect an intuition of causal relevance~\cite{yablo2003causal}. That is, the explanation should answer a question of sort \emph{when a GNN yields an outcome, which part of the input graph is causally relevant?} 

Another challenge is the so-called \emph{class imbalance}  which stems from the fact that fraudulent activities only correspond to a small fraction of the total activities within a financial system. 
As a result, only limited examples of fraudulent activities are available to learn from, which makes them more challenging to detect. Interestingly enough, Motie and Raahemi.~\cite{Motie2024} reveal in their review that research on GNN applications to financial fraud detection themselves carry an imbalanced rate of exploration; that is, supervised learning approaches make 88\% of all approaches, with semi-supervised and unsupervised approaches correspond to 9\% and 3\%, respectively. A semi-supervised approach here would mean learning a classifier from the labeled data points while learning structural properties from unlabeled data points. 

This paper aims to bridge the research gaps in applying GNN for anomaly detection in the financial domain by creating the first semi-supervised GNN model on the bipartite graph while simultaneously providing causal explanations for the identified anomalies. The unsupervised GraphBEAN model \cite{Fathony2023} is the most closely related work to ours in the context of anomaly detection, as it compresses a bipartite social media network input graph. However, it lacks explainability in its model outputs. Our work extends the GraphBEAN architecture to adapt it for real-world financial transaction datasets and incorporates Granger causal explanations for model outputs, which is crucial to use a deep learning model in the financial domain.

To highlight, we make the following contributions:

\begin{enumerate}
\item We propose a semi-supervised architecture, called SAGE-FIN, for financial fraud detection, and show its potential on generalizing beyond bi-partite graphs (i.e., $k$-partite graphs). 

\item We evaluate SAGE-FIN on a publicly available real-world dataset: Elliptic++.

\item We offer insights into financial anomalies identified by SAGE-FIN through the lens of Granger causality, aligning with the specific needs of the financial sector.
    
\end{enumerate}

The rest of the paper is organized as follows. Next section presents the problem definition and key notations; Section 3 illustrates our main contribution -- the SAGE-FIN architecture with the Granger causal explanation; Section 4 presents experimental setup, Elliptic++ dataset and the results; Section 5 is the discussion section, sharing gained insights and the limitations; Section 6 is the related work; Section 7 (Conclusion) closes the paper.  

\section{Background knowledge and Notation}
In this section, we formally define the problem definition, provide relevant background knowledge and introduce the key notations upon which our proposed SAGE-FIN model is based. 
\subsection{Problem definition}
\begin{figure}
\includegraphics[width=0.9\textwidth]{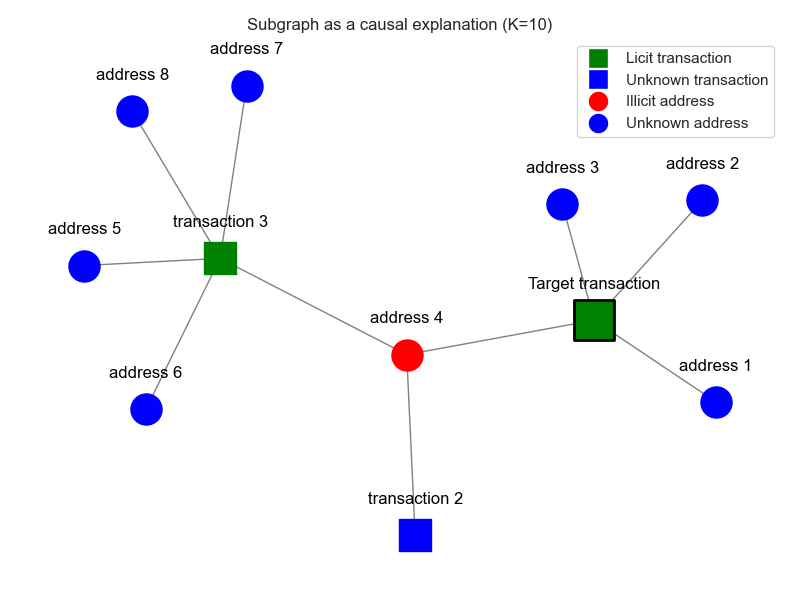}
\caption{An example graph of the Bitcoin wallets-transactions network.} \label{fig:example}
\end{figure}

In our scenario,  a monetary transfer between wallets is a transaction. This can be represented by a (bipartite) graph in which both wallets and transactions are represented by nodes; each having its own set of features. See Figure~\ref{fig:example} for an example sub-graph of the Bitcoin transaction dataset. More formally, we define a bipartite graph as $G=(V,U,E)$. For each node $v\in V$ and $u \in U$, a feature vector $x_v \in R^n$ and $x_u \in R^m$, sets $U$ and $V$ represent transactions and  wallets, respectively. Further, $e \in E$ represents an edge that connects a wallet to a transaction. For some nodes,  we have class labels $y\in \{0,1\}$. For example, in Figure~\ref{fig:example}, we have Wallet \textit{address 8} that sends some Bitcoin to Wallet \textit{address 4}. This transaction is shows as \textit{transaction 3}. Here \textit{address 4} is labeled as an fraudulent address and \textit{transaction 3} is labeled as a non-fraudulent transaction. The objective is to learn a model that predicts the classes of transactions and wallets, and provides an explanation for the predicted instance. 
\subsection{Message passing and graph convolution}
This research studies the heterogeneous bipartite graph structure which has different feature vectors for each node partition. This fills up the research gap that most of the literature worked on the monopartite graph whereas the financial transaction networks often exhibit the bipartite nature. To use all the feature vectors of both node types and edges, the message passing algorithm is designed such that it aggregates all information from the two node types and the edges. 

The SAGE-FIN network follows the GraphBEAN architecture proposed by \cite{Fathony2023} which comprises \(P\) (even number) convolution layers, where half of these layers are allocated for the encoder and the remaining half are utilized for the feature decoder. The \(q\)-th layer representation for node \(u_i\) within set \(U\), node \(v_j\) within set \(V\), and edge \(e_{i,j}\) within set \(E\) are presented as \(\mathbf{h}^{u{(q)}}_i\), \({h^{v(q)}_j}\), and \({h^{e(q)}_{i,j}}\) accordingly.

According to the GraphBEAN architecture proposed by \cite{Fathony2023}, to compute the new representations \({h^u_i}^{(q)}\), \({h^v_j}^{(q)}\), and \({h^e_{i,j}}^{(q)}\), the messages are collected and passed to the particular node and edge. The messages to a node \(u_i\) within set \(U\) come from the aggregator functions (average) over the neighboring node representations, the aggregator functions (average) over the edge representations connected to \(u_i\), and its own previous representations. A similar process is followed for the messages to node \(v_i\) within set \(V\). Messages are also collected and passed to edge \(e_{i,j}\) within set \(E\), which are simply derived from the nodes connected to the edge, namely \(u_i\) and \(v_j\) , and its own previous representation. The computations of messages passed to \(u_i\), \(v_j\), \(e_{i,j}\) are formulated as follows:

\noindent
\begin{equation}
\mathbf{h}^{u{(q)}}_i = \gamma_{\Theta} \left( \phi_{\Theta} (\mathbf{x}^u_i), \bigoplus_{v_j \in \mathcal{N}(u_i)} \phi_{\Theta} (\mathbf{x}^v_j), \bigoplus_{e_{i,j} \in \mathcal{M}(u_i)} \phi_{\Theta} (\mathbf{x}^e_{i,j}) \right) 
\label{eq:1}
\end{equation}

\noindent
\begin{equation}
\mathbf{h}^{v(q)}_i = \gamma_{\Theta} \left( \phi_{\Theta} (\mathbf{x}^v_j), \bigoplus_{u_i \in \mathcal{N}(v_j)} \phi_{\Theta} (\mathbf{x}^v_i), \bigoplus_{e_{i,j} \in \mathcal{M}(v_j)} \phi_{\Theta} (\mathbf{x}^e_{i,j}) \right) 
\label{eq:2}
\end{equation}

\noindent
\begin{equation}
\mathbf{h}^{e(q)}_{i,j} = \gamma_{\Theta} \left( \phi_{\Theta} (\mathbf{x}^u_j), \phi_{\Theta} (\mathbf{x}^v_j), \phi_{\Theta} (\mathbf{x}^e_{i,j}) \right) 
\label{eq:3}    
\end{equation}
\noindent where \(\phi_{\Theta}\) denotes the differentiable function to learn the features, \(\bigoplus\) represents the mean, max aggregation of the encoded features within each node partition and edge set, \(\gamma_{\Theta}\) concatenates the aggregated encoded feature vectors of both node partition and edge set together.

\begin{figure}
\includegraphics[width=\textwidth]{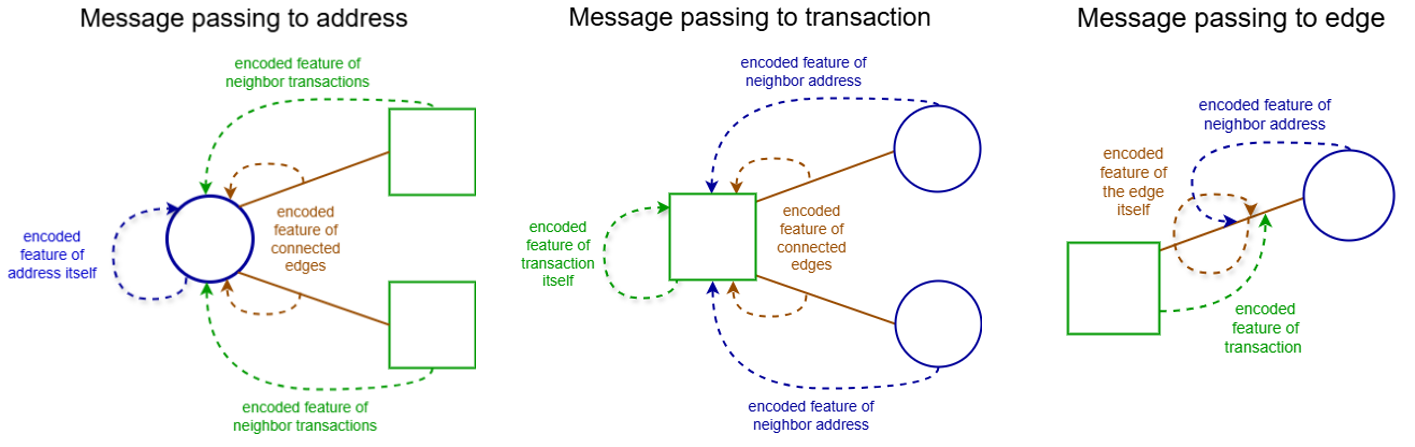}
\caption{Illustration of the message passing process for nodes and edges adapted from \cite{Fathony2023}.} \label{fig:messages}
\end{figure}

The message flows presented in the equations \ref{eq:1}, \ref{eq:2}, \ref{eq:3} are depicted in Figure \ref{fig:messages}. In each convolution layer, the message to the first node partition is aggregated from the information of this node in the previous layer, the information of the other node partition, and the information of the edges connecting these nodes. The same rule applies to the other node partition. For the edge, the message is passed from the information of that edge in the previous layers and the information of both node partitions on that edge.
After the messages are collected and passed to each node and edge, the subsequent representations are computed. The messages are passed to a linear operation and then the output of the linear operation is normalized using batch normalization. Subsequently, this normalized output is passed through a ReLU activation function.

\subsection{Granger Causality}
Granger causality, as originally proposed by Clive Granger \cite{Granger1969}, is a statistical hypothesis test for determining whether one time series is useful in forecasting another. Specifically, a time series $X_t$ is said to Granger-cause another time series $Y_t$ if past values of $X_t$ have a statistically significant effect on predicting future values of $Y_t$, given past values of $Y_t$ itself ~\cite{Granger1969}. Formally, $X_t$ Granger-causes $Y_t$ if:
\begin{equation}
\text{Var}(Y_{t+1} | Y_t, Y_{t-1}, ..., X_t, X_{t-1}, ...) < \text{Var}(Y_{t+1} | Y_t, Y_{t-1}, ...)   
\end{equation}
\noindent where:
\begin{itemize}
    \renewcommand{\labelitemi}{\textbullet}
    \item $\text{Var}(Y_{t+1} | Y_t, Y_{t-1}, \dots, X_t, X_{t-1}, \dots)$ is the variance of the prediction error of $Y_{t+1}$ using past values of both $Y$ and $X$.
    \item $\text{Var}(Y_{t+1} | Y_t, Y_{t-1}, \dots)$ is the variance of the prediction error of $Y_{t+1}$ using only past values of $Y$.
\end{itemize}

Within the context of this paper (i.e., Granger causality for Graph neural network), the Granger causal explainer algorithm generates a subgraph \(S\) for a node prediction from the original graph \(G\). The subgraph \(S\) would then contain relevant information to make a prediction (classification) of an instance if it is fraudulent or non-fraudulent. Fundamentally, the Granger causal explainer algorithm finds the edges in the graph \(G\) that significantly affect the outcome of the classifier (fraud or not) in the neural network. All these significant edges form the subgraph \(S\). The Granger causal explainer also implies that the outcome of the classifier does not change significantly when using the entire graph \(G\) or only a subgraph \(S\). The ultimate objective of the Granger causal explainer is to find a subgraph of the original network with a significant causal influence on the outcome of classifier in the graph neural network.

\section{Methodology}
\label{sec:methodology}
This SAGE-FIN model first employs the message passing algorithm that propagates also the edge attributes together with the node attributes to the graph convolution layers. The binary classification loss is added to the loss function. Here, the ground-truth labels of some data points guide the model to learn the anomalous patterns in conjunction with the node features and topological patterns of the entire graph, which includes a substantial amount of unlabeled data points. Building upon the SAGE-FIN model, a Granger causal explainer algorithm is implemented to extract the most critical nodes and edges. These elements significantly influence the classification of fraud versus non-fraud by the SAGE-FIN model, guided by the principle of Granger causality. From here onward, within this paper, we name Granger causal explanation as Causal explanation in short.

\subsection{Graph neural network architecture for anomalies detection}
\begin{figure}
\includegraphics[width=\textwidth]{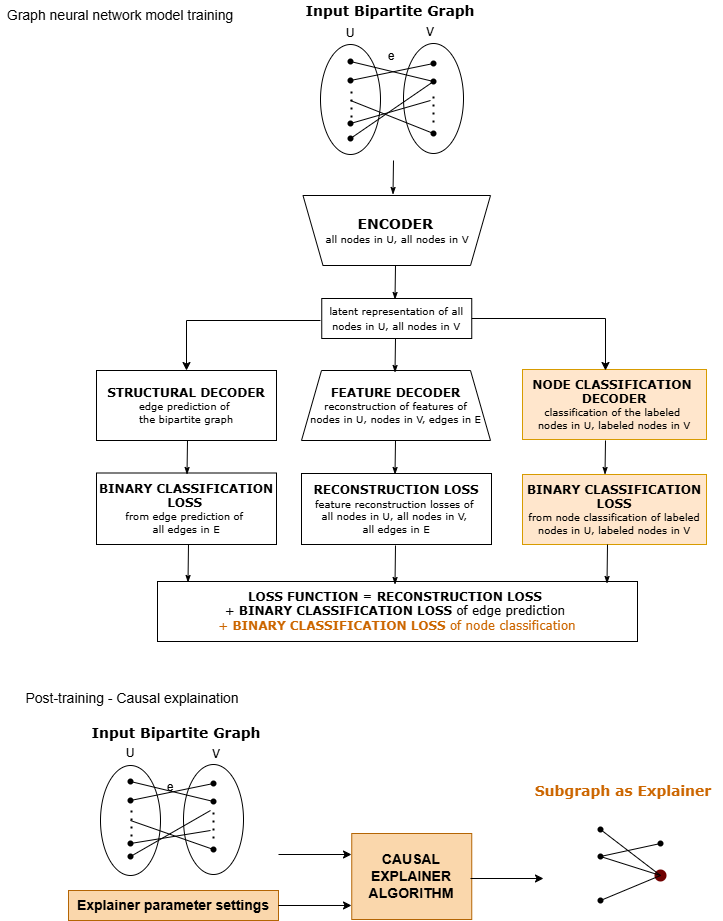}
\caption{The SAGE-FIN architecture adds a node classification decoder to the unsupervised GraphBEAN architecture \cite{Fathony2023} and supplement the neural network with Causal explanation layer after the model training. The new elements contributed by this paper are highlighted in orange.} \label{fig:SAGE-FIN}
\end{figure}

We propose a semi-supervised neural network architecture for anomaly detection in bipartite graphs with two main components: 1) unsupervised structure learning and 2) supervised classification learning. The first component consists of a (a) graph convolutional autoencoder that reconstructs the node's features and (b) an edge reconstruction module that predicts whether an edge exists between a pair of nodes. This aligns with the GraphBEAN model proposed by ~\cite{Fathony2023}. The second part, supervised learning, is the new element added to the GraphBEAN model. It is a linear classification head that classifies the node as fraudulent and non-fraudulent. The supervised part is only used for the limited data points with a label. This enhancement enables the architecture to function within a semi-supervised learning paradigm, as illustrated in Figure \ref{fig:SAGE-FIN}.
Adopted and modified from \cite{Fathony2023}, the training objective function is the combination of the reconstruction loss from the feature decoder, the binary classification loss of the structure decoder's edge prediction and the binary classification loss of the node classification decoder for the labeled nodes. 

\subsection{Causal explanation for anomaly detection outcomes}
After the model's classification stage, a Causal explainer algorithm extracts the nodes and edges that have the largest Granger causal contribution to the model's fraud/non-fraud classification. The distillation algorithm introduced by \cite{Lin2021} is modified to be compatible with the bipartite graph instead of the unipartite graph. Each specific node being classified by the SAGE-FIN model is examined. To quantify the influence of individual edges on the model output, the loss function produced by the model when operating on the entire graph \(G\) is compared with the loss function produced by the model when excluding a particular edge of the original graph. The Granger causal effect of that particular edge on the SAGE-FIN's output is then assessed by the reduction in the loss function, representing the change in loss function due to the removal of that edge. Formally,
\begin{equation}
    C_j = L_2 - L_1
\end{equation}
\noindent where,
\begin{itemize}
    \renewcommand{\labelitemi}{\textbullet}
    \item $C_j$ is the Granger causal effect of edge $e_j$
    \item $L_1$ is the loss function of the node classification decoder in SAGE-FIN for the entire graph
    \item $L_2$ is the loss function of the node classification decoder in SAGE-FIN when excluding edge $e_j$ from the original graph
\end{itemize}

Essentially, $C_j$ reflects the individual causal impact of the edge $e_j$ within the original graph \(G\) on the resulting classified nodes. By ordering the edges based on their causal contributions $C_j$, the top $K$ edges which constitutes a connected subgraph \(S\) signify the most influential connections and can be extracted to provide a causal explanation for the SAGE-FIN's predictions.

To identify a subgraph \(S\) that serves as a causal explainer for a node $Y$ labeled by the pre-trained SAGE-FIN model, an algorithm is employed. First, we define a set $N$ comprising all nodes within an $n$-hop neighborhood of $Y$. The parameter $n$ is determined based on the application-specific requirements, balancing the comprehensiveness of the subgraph \(S\) and the follow-up investigation complexity. Secondly, a list of edges of each node in the set $N$ is constructed based on the graph's adjacency matrix. After that, the causal explainer algorithm iterates over the neighborhood of $Y$, proceeding from the first to the $n^{th}$ hop. For each node encountered, and for each edge linked to this node, the algorithm calculates the loss function $L_2$ representing the loss when this edge is excluded and compares it with the initial loss function $L_1$ in the pre-trained SAGE-FIN model using the entire graph \(G\). If $L_2$ is bigger than $L_1$, it indicates a causal influence of this edge on the model's classification and this edge is appended to a list \(\mathcal{C}\). This process is repeated for all nodes within $n$-hop neighborhood. Finally, the algorithm selects the top K edges from the list \(\mathcal{C}\) with the most significant causal contribution $C_j$, ensuring that these edges form a connected subgraph containing node $Y$. This resulting subgraph \(S\) is then designated as the causal explainer for the target node $Y$. 

\section{Experimental Setup and Evaluation}

In this section, we describe the Elliptic++ dataset and outline the data pre-processing steps we performed.
\subsection{Dataset description and analysis}
Blockchain network offers a unique and accountable channel for financial forensics by mining its transparent and immutable transaction data. Recently, there has been a surge in the application of machine learning models trained with cryptocurrency transaction data for anomaly detection, such as money laundering and other fraudulent activities. For the task of anomaly detection using semi-supervised learning techniques within the GNN framework, the real-world Elliptic++ Bitcoin data is utilized. The dataset is presented in a bipartite format, where monetary flows from wallets to specific transactions are presented as edges, wallet addresses and transactions are depicted as two different node partitions. The address-transaction graph is visualized in Appendix \ref{sec:apx:A}. This representation facilitates the detection of fraudulent transactions and wallets addresses within the Bitcoin network. The source and detailed description of the data collection pipeline can be found in the paper \cite{Elmougy2023}. 

The Elliptic++ dataset resembles the heterogeneous graph structure of many real-world financial networks where each node partition has different features size. Within the Elliptic++ dataset, the Bitcoin transactions node set comprises 203,769 transactions (nodes), each characterized by 165 features. These features include 93 local features representing local information about the transaction (for example, the number of inputs/outputs, transaction fee, output volume) and 72 aggregated features derived by aggregating transaction information one-hop forward/backward \footnote{Due to intellectual property issues, the data source does not provide an exact description of all the features in the dataset}. The Bitcoin wallet addresses node set consists of 822,942 wallet addresses, each described by 56 features related to transactions, such as total Bitcoin transacted (sent and received), total fees in Bitcoin, and temporal features such as the number of blocks between transactions and the number of interactions among addresses. The Elliptic++ dataset is set up with high quality, ensuring no missing values within its comprehensive set of features. The main properties of the graph are shown in the Table \ref{table:eda}
\begin{table}
\centering
\caption{Key properties of Elliptic++ dataset}\label{table:eda}
\begin{tabular}{lcc}
\hline
 & Transaction set & Wallet set \\
\hline
\#Features  & 165 & 56 \\ 
\#Non-fraudulent nodes  & 42019 (21\%) & 251088 (31\%) \\ 
\#Fraudulent nodes & 4545 (2\%) & 14266 (2\%) \\ 
\#Unknown nodes & 157205 (77\%) & 557588 (67\%) \\
\hline
\end{tabular}
\end{table}
The bipartite graph, containing Bitcoin transaction and wallet node sets, spans over 49 timestamps. The interval between two timestamps is two weeks. The labels of the fraudulent/non-fraudulent for nodes are provided by the author \cite{Elmougy2023} based on the original Elliptic dataset published on Kaggle. The node features and labels are proprietary but publicly accessible.
\subsection{Data preprocessing and parameter settings}
\begin{figure}
\includegraphics[width=0.8\textwidth]{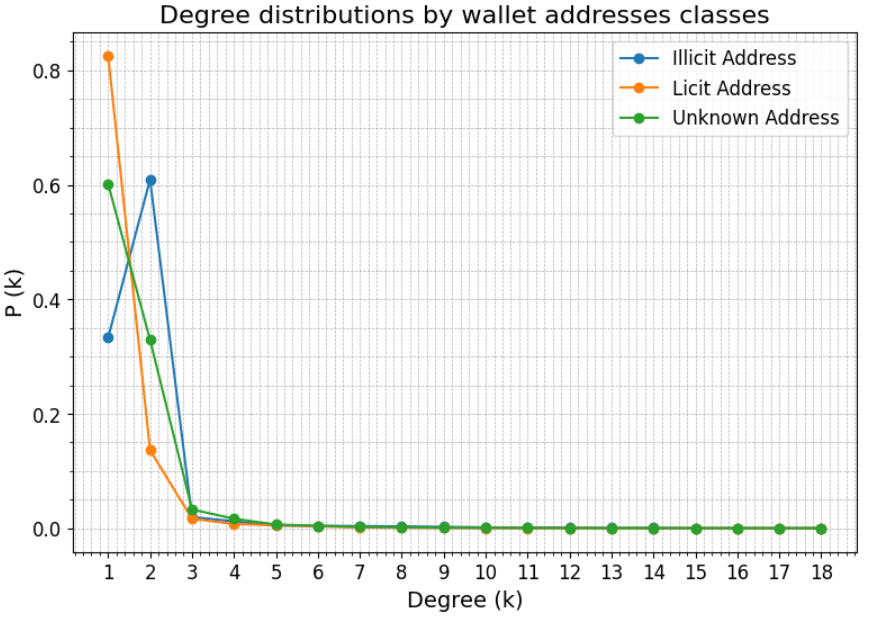}
\caption{Degree distribution of wallet address node set} \label{fig:P(k)_U}
\end{figure}

\begin{figure}
\includegraphics[width=0.8\textwidth]{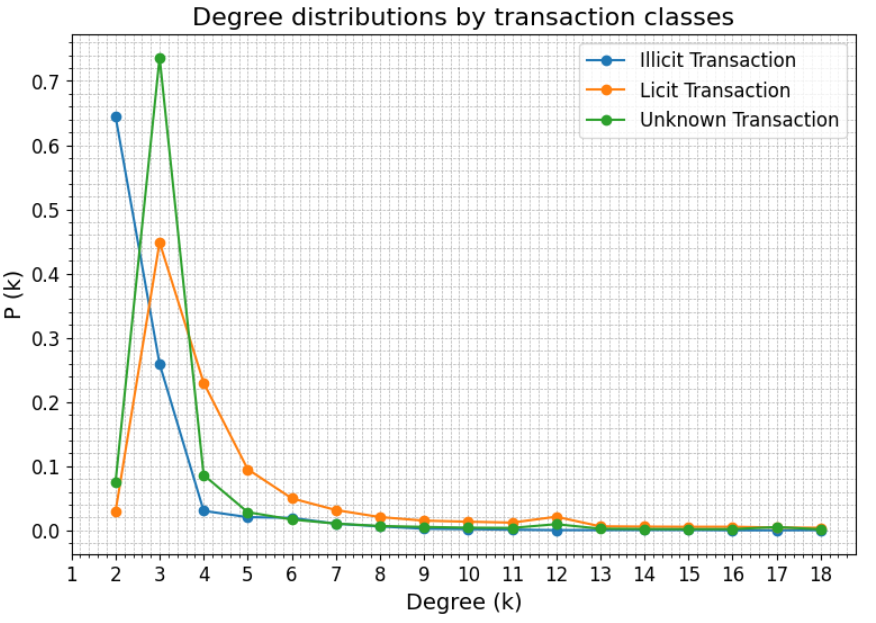}
\caption{Degree distribution of transaction node set} \label{fig:P(k)_V}
\end{figure}

First, the features and classes of wallet addresses and transactions are integrated with the addresses-transactions edge list to create a comprehensive list of nodes and edges for the graph construction. Among 165 transaction features, 72 aggregated features are removed from the model training because they are derived by aggregating transaction information one-hop forward/backward, which functions as the propagation of the message in the graph convolution layers. Therefore, only 93 local transaction features are merged into the wallets-transactions edge list. Subsequently, the dataset is divided into training, validation, and test sets using a random split with a ratio of 70\%-15\%-15\%. All features in the wallet addresses dataset are scaled using the standardization transformation.The normalized features of all nodes and edges are then used by the SAGE-FIN model.
The SAGE-FIN architecture consists of four layers: two convolution layers for the encoder and two convolution layers for the feature decoder. The structure decoder and node classification decoder include Multilayer Perceptrons with four dense layers. For training SAGE-FIN model, the Adam optimizer is used with a learning rate of 0.005. The dimension of latent variables and hidden layers is set to 32. Finally, for the edge prediction in the structure decoder, the negative cases (non-connected node pairs) are set to five times the number of edges (connected node pairs) in the graph. These hyperparameters were selected after several experiments and the combination of these values yielded optimal performance in the validation set. \textcolor{black}{The SAGE-FIN model is trained for 200 epochs.} The model performance is evaluated using the following metrics: precision, recall, and F1 score.
After the SAGE-FIN model identifies the anomalies, a causal explainer algorithm is implemented to extract the causality explanation for the GNN's classification. The goal is to identify a subgraph \(S\), determined by the number of hops ($n$-hop) and number of edges ($K$) from the target node, that significantly influences the output of the model. Since the ground-truth motif for the Elliptic++ dataset is unavailable, the values of $n$-hop and $K$ are selected qualitatively to ensure that the subgraph \(S\) is human-interpretable and suitable for further investigation. A $K$ value that is too small may yield insignificant explanations, while a $K$ value that is too large may result in an overly complex subgraph \(S\). From Figure \ref{fig:P(k)_U} and Figure \ref{fig:P(k)_V}, it can be observed that the majority of the wallet addresses and transactions have no more than five direct edges and a minimal number of addresses and transactions possess more than ten direct edges. Therefore, it has been determined that the nodes within a 4-hop radius from the target node are considered, and the causal explainer algorithm retrieves the $K$ (i.e., in our case $K=10$) edges that exhibit the most significant causal contribution to the classification of the target nodes.

\section{Results}
\label{sec:results}
Our evaluation consists of two parts:  anomaly detection and their Granger causal explanations.
\subsection{Anomaly detection}
To evaluate the anomaly detection of the SAGE-FIN model on the Elliptic++ dataset, metrics including precision, recall, and F1-score were employed to evaluate the anomaly detection on the labeled data with ground-truth annotations. The performance of the SAGE-FIN model stabilized after 40 epochs and achieved its optimal F1-score for the labeled nodes at epoch 190. The training set attained F1 scores of 0.908 and 0.918 for the wallet addresses and transactions, respectively. The validation set recorded F1 scores of 0.798 and 0.812 for the wallet addresses and transactions. With those settings, the F1 scores for the test set are 0.806 and 0.807.
The performance of the SAGE-FIN is benchmarked against the state-of-the-art (SOTA) machine learning models on the labeled dataset. We consider the following SOTA models: Logistic Regression (LR), Random Forest (RF), Multilayer Perceptrons (MLP), and Extreme Gradient Boosting (XGB), as examined in the study by \cite{Elmougy2023}. The classification performances on the labeled data by the SAGE-FIN model and the four state-of-art machine learning models are presented in Table \ref{table:wallet}:

\begin{table}
\centering
\caption{Fraudulent wallet and transaction detection results}\label{table:wallet}
\begin{tabular}{l|ccc|cccl}
&&Wallet &&&Transaction \\
\hline
Model & Precision & Recall & F1 score & Precision & Recall & F1 score \\
\hline
SAGE-FIN & 0.802 & 0.775 & 0.806 & 0.753 & 0.792 & 0.807 \\ 
LR & 0.491 & 0.049 & 0.089 & 0.649 & 0.091 & 0.159 \\  
RF & 0.968 & 0.793 & 0.872 & 0.986 & 0.829 & 0.899 \\ 
MLP & 0.823 & 0.414 & 0.551 & 0.850 & 0.856 & 0.853 \\ 
\text{XGB} & \textbf{0.958} & \textbf{0.826} & \textbf{0.887} & \textbf{0.974} & \textbf{0.872} & \textbf{0.920} \\  
\hline
\end{tabular}
\end{table}

For the unlabeled nodes, since ground-truth annotations are unavailable for evaluation, address-transaction edge prediction accuracy can be used as a proxy for the evaluation. Anomalies can be interpreted as nodes involved in unexpected edges, where interactions between particular wallet addresses and transactions should not occur. The F1-score convergences for edges in the training and test sets after 190 epochs. The training set achieved an F1-score of 0.908 for the address-transaction edge. The validation set recorded an F1-score of 0.874 for the address-transaction edge. Under those settings, the F1-score for the test set is 0.883. 

In summary, the SAGE-FIN model propagates the node's features through the neural network architectures and learns the graph topological structure, thereby enabling the detection of both node and edge anomalies.

\subsection{Causal explanation for anomalies}
\begin{figure}
\includegraphics[width=0.9\textwidth]{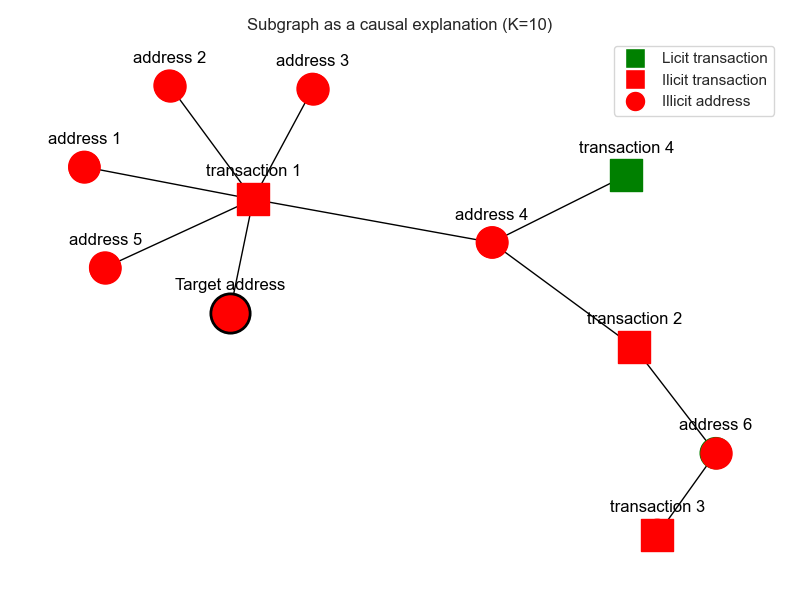}
\caption{Top 10 edges as the causal explanation for a target fraudulent wallet address (circles present addresses while squares present transactions)} \label{fig:fraud_top10}
\end{figure}

\begin{figure}
\includegraphics[width=0.9\textwidth]{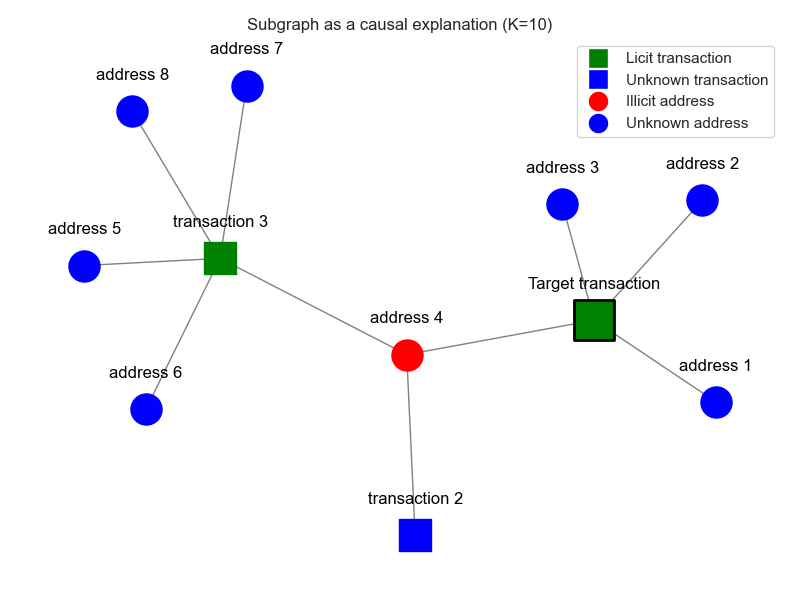}
\caption{Top 10 edges as the causal explanation for a target non-fraudulent transaction (circles present addresses while squares present transactions)} \label{fig:non_fraud_top10}
\end{figure}

Following the identification of node anomalies by the SAGE-FIN model, the subsequent objective is to extract causal explanations for the trained model's outcomes. Causal explanations aim to answer the question: when a graph neural network makes a prediction, which parts of the input graph are relevant? To verify the causal explanations derived from the causal explainer algorithm on the Elliptic++ dataset, given the detected anomalies, the following evaluation metrics are employed: 
\begin{enumerate}
    \item Within the test set containing labeled data, \(p(y|{S})\) should be close to \(p(y|{G})\). A good explainer should be able to generate more compact subgraphs yet maintain the prediction accuracy,
    \item The causal explanations should be visualized to allow qualitative performance analysis. 
\end{enumerate} 
Due to the substantial size of the dataset, this subsection demonstrates representative causal explanations for a detected anomalous wallet address in \ref{fig:fraud_top10} and a non-fraudulent transaction in \ref{fig:non_fraud_top10} for illustration purposes. The model's node classification decoder estimated a 0.84 probability that a particular wallet address is fraudulent and a 0.83 probability that a particular transaction is non-fraudulent. The causal explainer algorithm identified the top 10 edges that most significantly influenced the classifier's decision, as depicted in Figure \ref{fig:fraud_top10} for the fraudulent wallet address and Figure \ref{fig:non_fraud_top10} for the non-fraudulent transaction. By removing the other edges within a 4-hop radius from this identified fraudulent wallet address, the model's node classification decoder still estimated a 0.891 probability of fraud for this wallet address, close to the original probability of 0.84 without removing any edges. For the non-fraudulent transaction, the 10-edge subgraph led to a 0.91 probability of non-fraud, which is also close to the original probability of 0.83 without removing any edges. This proves that these top 10 edges primarily drove the classifier's decision from a casual perspective.

\section{Discussion}
\label{sec:discussion}
In this section, we reflect upon the evaluation of the performance of the SAGE-FIN model, the nature of explanations of the identified frauds, and some limitations. 
\subsection{On the performance of SAGE-FIN model}
We used the Elliptic++ dataset to detect fraudulent transactions and wallet addresses using the SAGE-FIN model. The edge prediction served as a proxy for detecting fraudulent nodes for unlabeled data. The model's performance for edge prediction across different datasets was compared. The model achieved an F1-score of 0.883 on the Elliptic++ test set, comparable to the original unsupervised GraphBEAN performance in \cite{Fathony2023} on the social media Wikipedia dataset, where the edge prediction F1-score was 0.902, demonstrating the stability of the model. For the labeled data, the semi-supervised version of the model demonstrated strong performance, with an F1-score of 0.806 for wallet addresses and 0.807 for transactions. The causal explainer algorithm extracts the subgraph \(S\) which is human-interpretable and suitable to investigate the fraudulent instance further.

The model is benchmarked against the SOTA machine learning algorithms, as shown in Table \ref{table:wallet}. Across different node partitions, the SOTA models themselves resulted in substantially better predictions for the transaction dataset in general, which has fewer nodes than the wallet address dataset. 
Moreover, our SAGE-FIN model integrates the features of wallet addresses and transactions within a single graph, thereby achieving comparable prediction performance for both node sets. 
Noteworthy is that GNN is also capable of handling larger datasets, whereas the performance of the SOTA methods deteriorates when the data size grows. Across several model choices, the model significantly outperformed the Logistic Regression model but did not surpass RF and XGB, primarily due to the latter model's noticeably higher precision. Overall, XGB is the best-performing model for fraud detection.
The Logistic Regression model, which assumes a linear relationship between the input features and the log probabilities of the output, often fails for complex data, particularly in cases involving complex relationships such as those in graph-structured data. 
In contrast, XGB builds the trees sequentially, with each new tree correcting errors from the previous ones. These approaches prevent overfitting and enhance effectiveness and accuracy on the Elliptic++ dataset. \textcolor{black}{There are also explainability methods for these SOTA models. Explainable Boosting Machine introduced by \cite{Caruana2015} combines cyclic gradient boosting with automatic interaction detection, resulting in models that are both accurate and inherently understandable. Similarly, SHAP (SHapley Additive exPlanations) developed by \cite{Lundberg2017} is a unified approach to interpreting model predictions, which has been applied to GBDTs to provide clear local explanations by attributing the contribution of each feature to the final prediction. However, while methods like SHAP and Explainable Boosting Machine provide clear, additive attributions for feature importance, these methods struggle to capture dependencies between features.} Overall, although the GNN did not perform as well as RF and XGB in anomalies detection, it can offer causal explanations for the identified fraud cases based on the graph structure which is the main advantage of our work. \textcolor{black}{They are specifically designed to identify critical subgraphs, nodes, and edges that influence predictions. These explainers go beyond traditional feature importance scores by uncovering relational patterns and hierarchical dependencies in graph data, which is crucial for domains like financial fraud detection.} This capability helps auditors to investigate the group associated with the identified fraud and its interactions, thereby providing a comprehensive understanding of the anomaly under investigation.

To evaluate the anomalies detection for unlabeled data, the address-transaction edge prediction accuracy can be used as a proxy. Anomalies can be interpreted as nodes involved in unexpected edges, where interactions between particular wallet addresses and transactions should not occur. \textcolor{black}{However, edge prediction typically achieves higher accuracy than node anomaly detection because it directly leverages structural patterns and historical transaction behaviors, making it easier to identify likely connections. In contrast, node anomaly detection requires distinguishing between normal and abnormal behaviors at the entity level, which is inherently more complex due to the diverse transaction patterns of different users. Moreover, anomalies often represent rare or novel cases that lack sufficient historical data for reliable classification, whereas edge prediction benefits from abundant observed relationships that improve learning and generalization. Consequently, while edge prediction can serve as a useful proxy for node anomaly detection, its superior accuracy highlights the challenge of precisely identifying anomalous nodes solely based on their individual attributes and sparse interactions.}
\subsection{Causal explanation for the identified frauds}
\begin{figure} 
\includegraphics[width=0.9\textwidth]{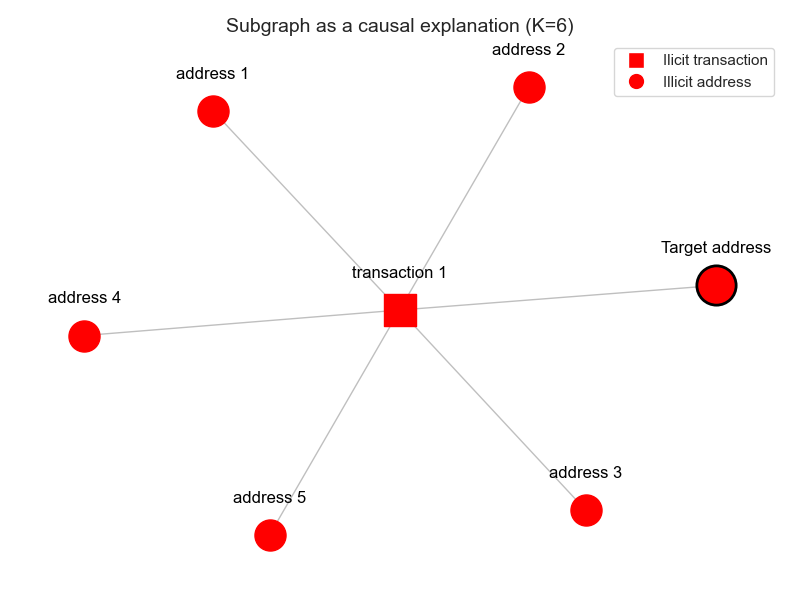}
\caption{Top 6 edges as the causal explanation for a target fraudulent wallet address (circles present addresses while squares present transactions)} \label{fig:fraud_top6}
\end{figure}

\begin{figure}
\includegraphics[width=0.9\textwidth]{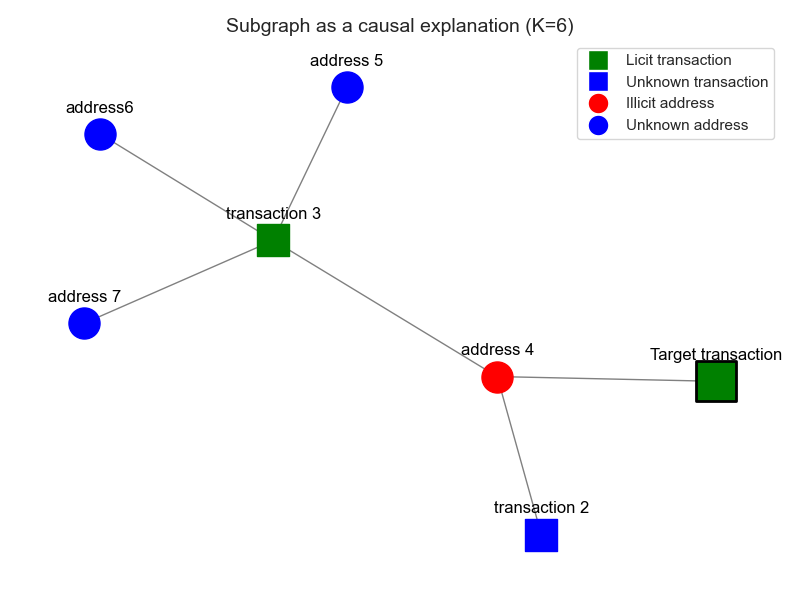}
\caption{Top 6 edges as the causal explanation for a target non-fraudulent transaction (circles present addresses while squares present transactions)} \label{fig:non_fraud_top6}
\end{figure}

The causal explanation for a representative fraudulent wallet address and a non-fraudulent transaction in the Elliptic++ dataset are illustrated in Figure \ref{fig:fraud_top10} and Figure \ref{fig:non_fraud_top10}.
The causal explainer algorithm effectively identifies the essential components — neighboring nodes — that contribute to the fraud label of a particular identified fraudulent activity. In Figure \ref{fig:fraud_top10}, the fraudulent activity (denoted by the red node) was determined by a fraudulent transaction (denoted by the red squares), which was associated with multiple fraudulent wallet addresses (denoted by red circles). One non-fraudulent transaction (the green square) was included in the subgraph because it connected to a fraudulent wallet address and contained relevant information for the model decision. On the other hand, in Figure \ref{fig:non_fraud_top10}, the non-fraudulent transaction (denoted as the green square) was decided by three unknown wallet addresses (denoted by blue circles) and a fraudulent address (the red circle). Even though the associated addresses were unknown, their features did not exhibit any suspicious patterns. Moreover, the fraudulent address played a role in the Granger causal explanation of the non-fraudulent transaction. Because this fraudulent address was linked to a non-fraudulent transaction, which suggests that the target transaction is classified based on the features of multiple nodes and edges in its neighborhood.   
For the sensitivity analysis, the subgraphs with ten edges for a fraudulent wallet and a non-fraudulent transaction were reduced further to smaller subgraphs with only six edges, as shown in Figure \ref{fig:fraud_top6} and Figure \ref{fig:non_fraud_top6}. This analysis reveals that the classification of the identified fraud was predominantly driven by a single fraudulent transaction connected to the other five non-fraudulent wallet addresses. Even after further edge removal, the model's node classification decoder estimated a probability of 0.96 for the fraud label of this wallet address. This probability was less aligned with the original probability of 0.84 when no edges were removed, compared to the 10-edge subgraph with an estimated probability of 0.891. For the non-fraudulent transaction, the classification was mainly driven by the cluster with some labeled nodes, leading to a probability of 0.90 for the non-fraud label of this transaction which is close to subgraph of size 10-edge with an estimated probability of 0.91. This emphasizes the importance of the labeled nodes in deciding whether a node is a fraud or not. Such finding also indicates that a subgraph with ten edges is likely to offer a more comprehensive view of the causal explanation for the GNN classification.

\subsection{Limitations}
In this study, the application of the SAGE-FIN neural network for detecting fraudulent activities within a financial dataset is explored. The model is designed to utilize the rich information inherent in the nodes as well as the overall structure of the graph. However,  the model neglected the graph's temporal dimension. The Elliptic++ dataset includes time step information, which could be crucial for understanding the evolution of fraudulent activities over time. \textcolor{black}{Incorporating temporal graph neural network such as Temporal Graph Attention Network or Dynamic Graph Neural Network could allow the model to capture temporal dependencies and evolving transaction patterns, potentially improving anomalies detection accuracy}. Moreover, looking at the explanatory subgraph can still be difficult to interpret to some extent in particular cases. On the causality side, it should be noted that unlike Pearlian causality \cite{pearl2016causal}, the notion of Granger causality is still based on correlation, and there can be undesirable hidden confounders which misguide the process of explaining actual causation. \textcolor{black}{Methods like structural causal models or causal representation learning could be used to identify confounding variables and enhance the causal interpretability of the anomalies detection explanation.}

\section{Related Work}
\label{sec:related_work}
We provide an overview of relevant literature, which consists of two aspects: GNNs for anomaly detection and causal explanations for GNNs.
\subsection{Graph neural networks for anomaly detection}
In many domains, the financial data forms a bipartite (or more generally $k$-partite) graph, such as bookkeeping in auditing, banking transaction and Bitcoin trading. Such bipartite graphs have distinct and rich node-and-edge attributes. Thus, the capability of an employed ML model to operate on bipartite node-and-edge-attributed graphs is crucial for real-world applications.

\subsubsection{
Supervised learning on real-world financial data} Bitcoin Elliptic dataset~\cite{Elmougy2023,Weber2019} is a widely used public dataset containing transactional data of the Bitcoin blockchain network. Graph Convolutional Networks are often used to detect fraudulent instances in graphs. Kipf et al.~\cite{Kipf2016} popularized the applications of such models in the machine learning domain. The methods can be categorized into spatial and spectral-based methods~\cite{Zhou2018,Zhang2019}. For example, \cite{Ghosh2023} employs metapath-based heterogeneous GNNs for collusion fraud detection, \cite{Jiang2023} predicts mutual fund returns with GNNs, \cite{Cao2023} proposes two neural networks to prevent financial contagion, \cite{Giudici2016,Billio2012,Battiston2012} apply financial networks to study contagion effects and systemic risks. However, prior studies on Elliptic rely on supervised learning methods, which is impractical in the financial industry.

\subsubsection{Semi-supervised learning on the uni-partite financial graphs} 
\cite{Jing2019} applied GraphSAGE for semi-supervised fraud detection in credit card transactions. \cite{Rao2021}  introduced a Knowledge-Guided Semi-Supervised GNN, using expert-designed rules to label data and train fraud detectors. \textcolor{black}{However, these models work only with unipartite graphs without capturing the interaction of the financial network. Financial data can be modeled as bipartite graphs with distinct features for each partition and rich node and edge attributes, which are essential for accurate anomaly detection and explainability. A model limited to uni-partite graphs would miss crucial information, making bipartite node-and-edge-attributed graph processing vital for such applications.}

\subsubsection{Unsupervised learning for bipartite non-financial graph}\cite{Fathony2023} introduced the unsupervised GraphBEAN model, which compresses a bipartite social network input graph into low-dimensional (node-only) latent representations, utilizing a series of customized graph convolution layers. Subsequently, the topological structure of the graph, as well as its node and edge attributes, are reconstructed. To improve the learning process on the unlabeled data, the domain knowledge can be used to first label the frauds, followed by a semi-supervised approach. 
\subsection{Causal explanation for graph neural networks}
At the intersection of XAI and Graph Neural Networks, we recognize two main categories: (a) instance-level explanations and (b) model-level explanations~\cite{Yuan2023}. The instance-level explanations depend on the input, such as gradient, perturbation, decomposition, and surrogate methods. In contrast, model-level explanations provide input-independent explanations, resulting in high-level understanding e.g., the XGNN model proposed by \cite{Yuan2020} which was designed to scrutinize graph patterns that lead to a specific class, the Bayesian graphic networks model by \cite{Giudici1996,Giudici2022} which takes into account the multivariate conditional dependencies among the nodes in the graph. We focus on \textbf{instance-level explanations} to explain why a particular instance is flagged as fraudulent. Examples include GNNExplainer \cite{Ying2019}, which offers a local explanation for an instance (a node/ an edge/ a graph) by identifying a compact subgraph that leads to its prediction, PGM-Explainer \cite{Vu2001} which models feature dependencies via conditional probability, Gem \cite{Lin2021} a model-agnostic method for various graph tasks, Shapley values \cite{Babaei2022} which provides the explanations for the predictions generated by a machine learning model. The Granger causality \cite{Granger1969} determines if one variable improves the prediction of another, providing transparent causal insights without relying on a specific model structure. 
On the other hand, function-based Causal discovery like Neural Granger Causality \cite{Tank2021} leverages deep learning to model complex, high-dimensional, and nonlinear relationships, which allows it to uncover more intricate causal dependencies. However, this comes at the cost of higher computational demands and reduced interpretability. While more flexible, its black-box nature limits its suitability in finance, where transparency is crucial for regulatory and policy decisions.
\section{Conclusion}
\label{sec:conclusion}
This paper proposed a semi-supervised model SAGE-FIN for anomaly detection in bipartite node-and-edge-attributed graphs, capable of detecting anomalies at both the node and the edge levels. The model leverages a small amount of labeled data to identify anomalies within a large volume of unlabeled data. The SAGE-FIN model has demonstrated commendable performance in terms of precision, recall, and F1-score metrics when applied to real-world datasets. It achieved F1 scores of 0.806, 0.807, and 0.883 for wallet addresses, transactions, and edges, respectively. While the SAGE-FIN model significantly outperformed the Logistic Regression model, it did not surpass the performance of Random Forest and XGBoost. However, the SAGE-FIN model's ability to provide (causal) explanations for identified fraud based on graph structure is a notable advantage. This capability has the potential to help auditors to  investigate the implicated group and its interactions thoroughly, offering a rich enough medium to pinpoint the reason for anomalies.

Additionally, the research introduced a Granger causal explanations for the SAGE-FIN model identified frauds cases. This is essential in domains like finance to ensure accountability in decision-making process.

Based on the identified limitations of the SAGE-FIN model as discussed in Section \ref{sec:discussion}, we recommend to incorporate the temporal dimension to develop a dynamic graph anomaly detection algorithm, and potentially extend it to go beyond Granger causality (i.e., Pearlian causality \cite{pearl2016causal}). 

\subsubsection*{Disclosure of Interests.}
The authors report no conflicts of interest, and declare that they have no relevant or material financial interests related to the research in this paper. The authors alone are responsible for the content and writing of the paper, and the views expressed here are their personal views and do not necessarily reflect the position of their employer.

\appendix
\section{Appendix A}
\label{sec:apx:A}
The Elliptic++ dataset presents a bipartite address-transaction graph that illustrates the flow of Bitcoin across transactions and addresses. This structure enables the evaluation of transaction purposes and the relationships among addresses within the same transaction: 
\begin{figure} 
\includegraphics[width=0.9\textwidth]{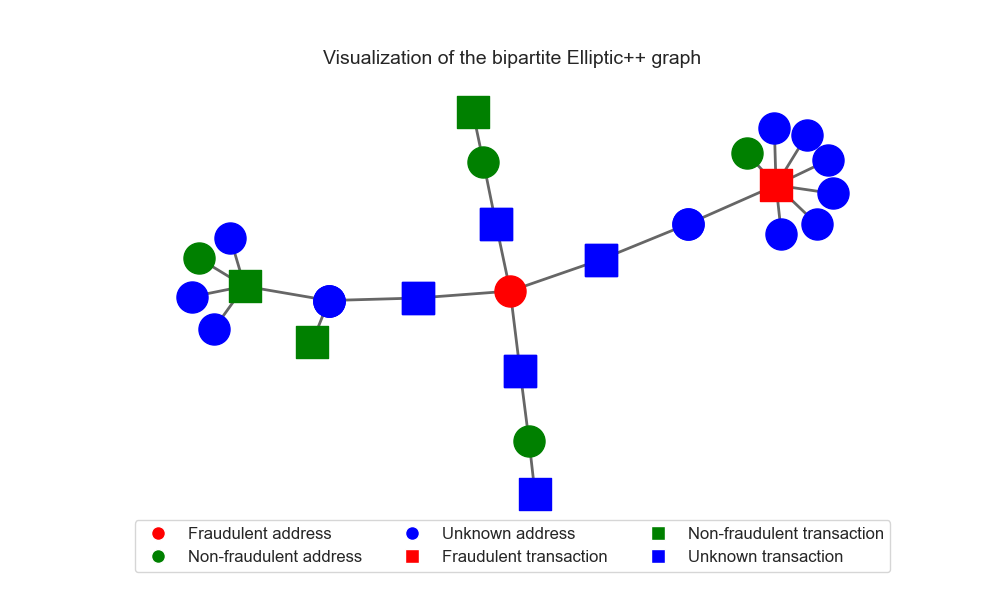}
\caption{Illustration of Bitcoin wallet addresses-transactions network as a bipartite graph} 
\end{figure}

\end{document}